%% file: main.tex
\documentclass{article}
\usepackage{spconf,amsmath,graphicx}
\usepackage{booktabs}
\usepackage{amsmath}
\newcommand\norm[1]{\left\lVert#1\right\rVert}
\usepackage{booktabs}
\usepackage{hyperref}

\title{UNIFIED HYPERSPHERE EMBEDDING FOR SPEAKER RECOGNITION}
\name{Mahdi Hajibabaei, Dengxin Dai}
\address{Computer Vision Lab, D-ITET, ETH Zurich}
\begin{document}
%
\maketitle

\begin{abstract}
Incremental improvements in accuracy of Convolutional Neural Networks are usually achieved through use of deeper and more complex models trained on larger datasets. However, enlarging dataset and models increases the computation and storage costs and cannot be done indefinitely. In this work, we seek to improve the identification and verification accuracy of a text-independent speaker recognition system without use of extra data or deeper and more complex models by augmenting the training and testing data, finding the optimal dimensionality of embedding space and use of more discriminative loss functions.\par 
Results of experiments\footnote{ \href{https://github.com/MahdiHajibabaei/unified-embedding}{https://github.com/MahdiHajibabaei/unified-embedding}} on \textit{VoxCeleb} dataset suggest that: (i) Simple repetition and random time-reversion of utterances can reduce prediction errors by up to 18\%. (ii) Lower dimensional embeddings are more suitable for verification. (iii) Use of proposed logistic margin loss function leads to unified embeddings with state-of-the-art identification and competitive verification accuracies.\par
\end{abstract}
\section{Introduction}
\label{sec:intro}

Speaker recognition is an area of research with more than 50 years of history and applications ranging from forensics and security to human-computer interaction in consumer electronics. Speaker recognition can be categorized into two tasks of text-dependent and text-independent speaker recognition with regard to the similarity of the uttered content between utterances. Text-independent speaker recognition task is the most general and non-trivial of the both that if performed accurately, can be used in everyday situations. Moreover, speaker recognition can be classified into two tasks of identification and verification. In identification, an utterance from one of the speakers within our training set will be given and the system needs to identify which speaker the utterance belongs to. In verification, two utterances from speakers not within our training corpora will be given and the predictor needs to decide whether these two utterances come from the same person and with what probability. The process of training a speaker recognition embedding, in both verification and identification, can be summarized as finding a functional mapping into a space in which utterances of the same speaker are embedded as close to each other as possible and as far away as possible from utterances of the other speakers.\par
Conventionally, utterances with various lengths, content and amount of environmental noise were transformed into variable number of vectors (or points) in a space spanned by features extracted from fixed length frames. The resulting vectors were then pooled together to evaluate a mean, mixture model or so called Supervectors. The elaborate choice of features depended on the amount of environmental noise, variations in recording setups, level of cooperation expected from speakers and etc. However, with immense increase in the amount of data that can be collected from the Internet and computation power provided by GPUs, optimality of conventional feature extraction and speaker modelling methods are under question.\par  
However, availability of datasets for speaker recognition with ever increasing sizes cannot be taken for granted. Until recently that \textit{Nagrani et al.} \cite{DBLP:journals/corr/NagraniCZ17} collected a dataset with more than two weeks of interviews in English with 1,251 celebrities, there was a lack of large publicly available dataset needed to train a deep model. Even though, the recently released \textit{VoxCeleb2} \cite{Chung18a} dataset includes more than 100 days of recordings from almost 6,000 speakers which is large enough to train a 50 layer Residual Network \cite{DBLP:journals/corr/HeZRS15}, methods improving the prediction accuracy without requiring more data or more parameters to be tuned can be used complementary to known methods to improve the prediction accuracies even further.
\par
\section{Related Works}
\label{sec:related}
The practice of text-independent speaker recognition has a longstanding history that falls outside the scope of this work and we advise the interested reader to consult literature review in this field \cite{KINNUNEN201012} for more information about traditional practices. Breakthroughs in speaker recognition due to use of Artificial Neural Networks dates back to more than a decade ago when they were used to jointly model and discriminate different speakers \cite{KINNUNEN201012}. However, the use of Convolutional Neural Networks developed for visual objects recognition in text-independent speaker recognition is a relatively new practice. \textit{Li et al.} \cite{li2017deep} trained a 28 layer ResNet on datasets with varying sizes for text-dependent and text-independent speaker recognition and found out that: training on larger datasets improves the prediction accuracy, the Residual Networks outperform stacked Gated Recurrent Units (GRU) with equal number of parameters and fine-tuning CNNs with more discriminative loss functions such as triplet loss significantly improves the verification accuracies. \textit{Nagrani et al.} \cite{DBLP:journals/corr/NagraniCZ17} showed that the relatively shallow VGG-M CNN trained on \textit{VoxCeleb} dataset can have verification and identification accuracies higher than that of all traditional models including GMM-UBM and i-vectors+PLDA. However, number of pairs that can be created for training an embedding using center loss, used by \textit{Nagrani et al.}, grows quadratically with the size of dataset and elaborate pair selection heuristics are needed to make the training on large datasets feasible. \textit{Cai et al.} \cite{cai2018exploring}  suggested to use the angular softmax loss \cite{liu2017sphereface}, that does not require pair selection, to create more discriminative embeddings and gained modest improvement in verification accuracy. Even though angular softmax loss creates embeddings with superior verification accuracy, the use of such discriminative loss functions has not resulted in identification accuracies superior to that of typical softmax loss.\par
In this paper, we first review the pipeline of the speaker recognition system that we used for all experiments in Section \ref{sec:pipeline}. Details about evaluation and training process along the dataset used in these two stages are presented in Section \ref{sec:experiments}. Effective augmentation methods in the context of CNNs that don't rely on external datasets are examined in Section \ref{sec:augmentation}. Effect of dimensionality of embeddings on verification and identification accuracies will be studied in Section \ref{dimensionalityResult} and different loss functions will be compared with regard to their resulting identification and verification accuracies in Section \ref{sec:lossFunction}. In addition effect of dropout \cite{srivastava2014dropout} on generalization power of trained models would be briefly analyzed in Section \ref{sec:dropout}. Our pipeline will be compared to other implementations of speaker recognition using \textit{VoxCeleb} in Section \ref{sec:discussion}. Finally, conclusions are presented in Section \ref{sec:conclusions}.

\section{Overview of pipeline}
\label{sec:pipeline}

\subsection{Feature Extraction}
Similar to \cite{DBLP:journals/corr/NagraniCZ17}, spectrograms are generated in sliding window fashion from hamming window of width 25 ms and step 10 ms. But in order to accelerate the training process, we take 512 element FFT of signal. The amplitude of FFT components plus the DC component of each frame are appended together to create a short-time Fourier transform (STFT) of size 300 $\times$ 257 (Temporal$\times$Spectral) out of every 3.015 second crop. Each frequency bucket would be normalized to have zero mean and unit variance. No voice activity detection (VAD) or augmentation through addition of noise or simulation of room impulse (as in \cite{okabe2018attentive}) is applied.\par
However, when 3 seconds crops are taken from an utterance, the starting point of these crop can be up to 3 seconds before the end of utterance, otherwise we have to extend the signal with a constant value or a sequence. We propose to extend the utterance by repeating it, so that the cropped signal can start at any point of utterance. In addition, we argue that time-reversed utterance still sound like it's uttered by the same person. The diverse embeddings yielded by feeding the randomly time-reversed crops of extended signal can be used to evaluate the models' parameters more accurately in training stage and yield better embeddings in testing stage.\par
\subsection{Convolutional Neural Network}
We chose the ResNet-20 architecture (shown in Table \ref{tab:networkStructure}) for all experiments in this work due to its low parameter count and short training time to yield a given accuracy. Moreover, the analyzed discriminative loss functions such as angular softmax loss \cite{liu2017sphereface} and additive margin softmax Loss \cite{wang2018additive} are all implemented in this architecture.

\begin{table}[t!]
  \centering
  \caption{ResNet-20 architecture used in this work}
  \label{tab:networkStructure}
  \begin{tabular}{ccc}
    \toprule
    Block Name &  Block's Structure &    Output Size \\
    \midrule
        \begin{tabular}{@{}c@{}} conv1\_1 \\conv1\_2\\conv1\_3\end{tabular} & 
        $\left[ \begin{tabular}{@{}c@{}} $3\times3$, 64 \\ $3\times3$, 64 \\ $3\times3$, 64 \end{tabular} \right]$ &   $150\times 129$ \\
    \midrule
        \begin{tabular}{@{}c@{}} conv2\_1 \\conv2\_2\\conv2\_3\end{tabular} & 
        $\left[ \begin{tabular}{@{}c@{}} $3 \times 3$, 128 \\$3\times 3$, 128\\$3\times 3$, 128\end{tabular} \right]$ & $75\times 65$ \\
    \midrule
        \begin{tabular}{@{}c@{}} conv2\_4 \\conv2\_5\end{tabular} & 
        $\left[ \begin{tabular}{@{}c@{}}$3\times 3$, 128 \\$3\times 3$, 128\end{tabular} \right]$ & $75\times 65$ \\
    \midrule
        \begin{tabular}{@{}c@{}} conv3\_1 \\conv3\_2\\conv3\_3\end{tabular} & 
        $\left[ \begin{tabular}{@{}c@{}} $3 \times 3$, 256 \\$3\times 3$, 256\\$3\times 3$, 256\end{tabular} \right]$ & $38\times 33$ \\
    \midrule
        \begin{tabular}{@{}c@{}} conv3\_4 \\conv3\_5\end{tabular} & 
        $\left[ \begin{tabular}{@{}c@{}}$3\times 3$, 256 \\$3\times 3$, 256\end{tabular} \right]$ & $38\times 33$ \\
    \midrule
        \begin{tabular}{@{}c@{}} conv3\_6 \\conv3\_7\end{tabular} &
        $\left[ \begin{tabular}{@{}c@{}}$3\times 3$, 256 \\$3\times 3$, 256\end{tabular} \right]$ & $38\times 33$ \\
    \midrule
        \begin{tabular}{@{}c@{}} conv3\_8 \\conv3\_9\end{tabular} & 
        $\left[ \begin{tabular}{@{}c@{}}$3\times 3$, 256 \\$3\times 3$, 256\end{tabular} \right]$ & $38\times 33$ \\
    \midrule
        \begin{tabular}{@{}c@{}} conv4\_1 \\conv4\_2\\conv4\_3\end{tabular} & 
        $\left[ \begin{tabular}{@{}c@{}} $3 \times 3$, 512 \\$3\times 3$, 512\\$3\times 3$, 512\end{tabular} \right]$ & $19\times 17$ \\
    \midrule
    pool1 & $19 \times 1$ &  $1\times 17$\\
    \bottomrule
    fc5 & 512 &  $1 \times 1$\\
    \bottomrule
  \end{tabular}
\end{table}

\subsection{Classification and Loss function}
A fully connected layer whose number of output is equal to number of identities in training set will be trained on top of fc5 layer and its output would be fed to the following loss functions:
\subsubsection{Softmax} The typical loss function used for training CNNs in visual object recognition. The value of the loss function is equal to negative log likelihood of predicting the true identity $(y_i)$: 
\begin{equation}
L_i=-\log({p_y}_i)
\end{equation}
\subsubsection{Angular Softmax (A-Softmax)} Dictates the angles between each sample and its ground truth class center to be $m$ times smaller than that of wrong classes'. In order to ease the training, the score given by A-Softmax is weight averaged with the typical cosine similarity with weight ($\lambda$) that decreases during training process. The best results are usually yielded with $m=4$ and $\lambda=5$ \cite{liu2017sphereface}.

\begin{equation}{f_y}_i=\frac{\lambda \norm{x_i} \cos({{\theta_y}_i})+\norm{x_i} \psi({{\theta_y}_i})}{\lambda+1}\end{equation}
\begin{equation}L_i=-\log\frac{e^{{f_y}_i}}{e^{{f_y}_i}+\sum_{j!=y_i}^{}{e^{\norm{x_i}\cos({\theta_j})}}}\end{equation}

\subsubsection{Additive Margin Softmax (AM-Softmax)}Forces the cosine similarity between sample and its true class to be $m$ more than similarity between sample and wrong classes. Also multiplies this difference by a scale parameter and feeds this to softmax with cross entropy loss function. The author failed to train scale ($s$), because after introduction of margin ($m$), $s$ remained constant and stopped increasing. Best results in face verification were achieved by setting $m$ to $0.35 - 0.4$ and $s$ to about $30$ \cite{wang2018additive}.\par
\begin{equation}L_i=-\log\frac{e^{s(\cos{{\theta_y}_i}-m)}}{e^{s(\cos{{\theta_y}_i}-m)}+\sum_{j!=y_i}^{}{e^{s.\cos({\theta_j})}}}\end{equation}

\subsubsection{Logistic Margin} We argue that instead of enforcing a geometrical margin there should be a probabilistic margin in the way that true class should be predicted with probability $e^\alpha$ times higher than that of wrong classes. The difference between this loss and AM-Softmax is that $s.m$ is treated as an independent variable subtracted from score of true class prior to feeding the logistic predictions to softmax with cross entropy loss. Non-normalized weight vector and an additional bias can be learned separately for each class. \par
\begin{equation}\label{eq:score} S_j=W_j\frac{x_i}{\norm{x_i}}+c_j\end{equation}
\begin{equation}L_i=-\log\frac{\exp{({S_y}_i-\alpha)}}{\exp{({S_y}_i-\alpha)}+\sum_{j!=y_i}^{}\exp{(S_j)}}\end{equation}
\section{Experimental Setup}
\label{sec:experiments}
\subsection{Dataset}
We chose \textit{VoxCeleb} dataset because of its moderate size (about two weeks of recordings), accurate labels and inclusion of fair amount of environmental noise. Moreover, there are recent speaker recognition systems trained on this dataset that we can use for comparison. The statistics of this dataset are given in Table \ref{tab:VoxCeleb}.

\begin{table}[h!]\label{VoxCeleb}
  \centering
  \caption{The statistics of the VoxCeleb dataset}
  \label{tab:VoxCeleb}
  \begin{tabular}{cccc}
    \toprule
    Task & Train Size/IDs & Val. Size/IDs   & Test Size/IDs \\
    \midrule
    Ident.	& 138,327/1,251	& 6,904/1,251	& 8,251/1,251 \\
    Verif.	& 141,940/1,211	& 6,670/1,211	&  4,872/40 \\

    \bottomrule
  \end{tabular}
\end{table}
\subsection{Training}
All CNNs in our experiments are trained on a single Titan Xp with batch size of 50, weight decay of 0.0005 and momentum of 0.93. Every training is done with SGD and the highest initial learning rate that doesn't result in divergence to avoid getting stuck in local minima and to accelerate the training. In order to avoid the occasional divergence during training, instead of multiplying the learning rate by 0.1 at each step we take 8 steps of multiplying by 0.75. In order to make sure that we train models for long enough, we doubled the number of iterations per step for some trainings, but we observed no improvement.\par

\subsection{Evaluation}
As it can be seen in Table \ref{tab:networkStructure}, in order to create an embedding in output of network, 19 rows of \textit{conv4-3}'s output are averaged and then fed to \textit{fc5}. To evaluate an embedding for utterances with different lengths, we take 50  random 3 second crop of signal for identification and verification, feed them to network and average resulting embeddings. Even though this methods is not the best pooling method possible \cite{Chung18a,cai2018exploring,okabe2018attentive}, resulting accuracies are not expected to change comparatively. For the task of identification, we report the Top-1 and Top-5 accuracy and for verification we report the equal error rate (EER) and minimum of detection cost function ($C_{det}$ in Equation \ref{eq:cDet}) on pairs suggested by \textit{Nagrani et al.} The detection cost is evaluate for $C_{miss}=C_{fa}=1$ and $P_{tar}=0.01$.\par
 \begin{equation}\label{eq:cDet}
 C_{det}=C_{miss} \times P_{miss} \times P_{tar} + C_{fa} \times P_{fa} \times (1- P_{tar})
 \end{equation}
 Note that variations of about 0.1\% in identification accuracies and 0.01 in $C^{min}_{det}$ are expected between models trained with the exact same settings, thus drawing conclusions based on small differences is not recommended.\par
\section{Experimental Results}
\label{sec:result}
\subsection{Augmentation}\label{sec:augmentation}
 To see whether mentioned augmentation techniques improve the prediction accuracy, we train and evaluate our CNN with and without these techniques and compare their prediction accuracies.\par

\begin{table}[h!]\label{augmentationResult}
  \centering
  \caption{Effect of augmentation on prediction accuracies}
  \label{tab:augmentationResult}
  \begin{tabular}{ccccc}
    \toprule
    Stage  & Top-1 (\%)   & Top-5 (\%)   &  EER (\%) & $C^{min}_{det}$ \\
    \midrule
    None& 87.5  & 96.0 & 7.55 & 0.609\\
    Testing& 89.2 & 96.3 & 7.14  &0.566\\
    Training&87.8   &95.9   &7.60   &0.621\\
    Both &89.7   &96.7   &6.98   &0.572\\
    \bottomrule
  \end{tabular}
\end{table}
As it can be seen in Table \ref{tab:augmentationResult}, applying the augmentation in both training and testing stages can reduce the identification error by more than 17\% and results in noticeable improvement in verification accuracies. In result, we will apply this augmentation during training and evaluation of all models used after this experiment.\par 
\subsection{Dimensionality of Embeddings}
Similar to \textit{Schroff et al.} \cite{schroff2015facenet} that investigated the effect of dimensionality of face embeddings on prediction accuracies, we train models that project utterances to embeddings spaces with different dimensionalities to see if there exists an optimal value for embeddings' dimensionality. The loss function used for training CNNs in this section is the typical Softmax with cross entropy loss function. As it can be seen in Table \ref{tab:dimensionalityResult}, dimensionality of embedding has a noticeable effect on prediction accuracies and lower dimensional embeddings are more suitable for verification while 256 seems to be the best dimensionality for identification.\par

\begin{table}[h!]\label{dimensionalityResult}
  \centering
  \caption{Effect of embedding dimensionality on prediction accuracies}
  \label{tab:dimensionalityResult}
  \begin{tabular}{ccccc}
    \toprule
    \#dim  & Top-1 (\%)   & Top-5 (\%)   &  EER (\%) & $C^{min}_{det}$ \\
    \midrule
    512& 89.7 & 96.7 & 6.98 & 0.572 \\
    256& 89.9 & 96.8 & 6.78 & 0.578 \\
    128& 89.9 & 96.7 & 6.50 & 0.556 \\
    64 & 88.6 & 96.6 & 6.27 & 0.537 \\
    \bottomrule
  \end{tabular}
\end{table}
\subsection{Loss Functions}\label{sec:lossFunction}
Typically, prior to training networks with softmax and cross entropy loss function, the coefficients are initialized with numbers randomly sampled from Gaussian distributions or by using methods such as \textit{Xavier} \cite{glorot2010understanding}. However, initializing the networks' coefficients prior to training them with more discriminative loss functions with aforementioned methods often results in convergence to sub-optimal solutions \cite{DBLP:journals/corr/LiMJLZLCKZ17} or no convergence at all. Similar to \cite{DBLP:journals/corr/NagraniCZ17,DBLP:journals/corr/LiMJLZLCKZ17}, we first train our models with softmax and cross entropy loss function and use the estimated coefficients for training more discriminative losses. Details on training procedure used for each loss function are as follows:
\subsubsection{Softmax}
Training is done with the initialization method of original ResNet-20. Training starts with learning rate of 0.05 and finishes after 22 steps of 2800 iterations. The validation error stops decreasing after 20 steps but we keep training for extra 2 steps anyway.
\subsubsection{A-Softmax}
Since using the coefficients of the \textit{fc5} layer of softmax trained network resulted in divergence during training, we initialized the coefficients of \textit{fc5} using \textit{Xavier} and layers prior to \textit{fc5} were initialized with softmax trained networks ($dim=64$). Initial learning rate is 0.0133 and training is done for 20 steps of 2000 iterations. Furthermore, \textit{gamma} is set to 0.015 and $\lambda_{min}$ to 5. The models with 256 and 512 dimensional embeddings did not converge.\par
\subsubsection{AM-Softmax}
Training starts with coefficients of the CNN trained with typical Softmax loss (with the same embedding dimensionality) and learning rate of 0.005 and ends after 15 steps of 4000 iterations. In line with \cite{wang2018additive}, we chose a constant value for scale of all classes and margin ($s=50$ \& $m=0.4$) and fine-tuned the rest of network.\par
\subsubsection{Logistic Margin}
The training conditions are same as \textit{AM-Softmax} but a weight vector and bias are created by a \textit{Scale} layer with bias on top of fully connected layer of \textit{AM-Softmax} whose output is passed to \textit{Label Specific Add} layer that adds -25 to the score of true class.

\subsection{Dropout}\label{sec:dropout}
Dropout \cite{srivastava2014dropout}, is a method known to be effective in reducing the variance of parameter estimation and improving the generalization power and accuracy of the trained models. However, in original ResNet \cite{DBLP:journals/corr/HeZRS15} implementation, no dropout was used. In this part of experiments, we apply dropout with probability of 50\% before feeding the output of pooling to \textit{fc5} to see whether it will result in any improvement in identification and/or verification accuracies.\par  

\begin{table*}[t!]
\centering
\caption{Prediction accuracies of models trained with different loss functions and embedding dimensionalities, w/wo dropout}
\resizebox{\linewidth}{!}
{\input{results.tex}}
\label{tab:results}
\end{table*}
\section{Discussion}
\label{sec:discussion}
As it can be seen in Table \ref{tab:results}, logistic margin loss without application of dropout performed the best in identification regardless of dimensionality. AM-Softmax trained with dropout outperformed other combinations of loss with or without dropout in verification regardless of dimensionality. Except for softmax loss, applying dropout resulted in improvement in verification accuracy. Decent verification accuracy of the A-Softmax trained embeddings came at the cost of identification accuracy inferior to embeddings trained with any other loss functions. Softmax loss function performed inferior to AM-Softmax and logistic margin in both identification and verification tasks. To the best of our knowledge, there is no implementation or model outperforming the 512 dimensional embedding trained by logistic margin loss in identification.\par
Accuracies of our models compared to others' can be seen in Table \ref{tab:comparativeResult}. In the upper section of the table all models trained on \textit{VoxCeleb} that utilized simple average pooling are compared together. The middle section is devoted to models trained on \textit{VoxCeleb} that utilize some type of attentive pooling such as Learnable Dictionary Encoding \cite{cai2018exploring} or Attentive Statistics \cite{okabe2018attentive}. The verification accuracies of models trained on much larger and more diverse \textit{VoxCeleb2} with different attention mechanism is shown in bottom of Table \ref{tab:comparativeResult}. \par 
In order to have a fair comparison we have to consider that \textit{Okabe et al.} \cite{okabe2018attentive} increased the robustness of their network through augmenting the training data with large number of background noise and room impulse responses from PRISM \cite{ferrer2011promoting} corpora and by using this corpora even i-vector+PLDA implementation had accuracy superior to that of CNNs trained by \textit{Nagrani et al.} \textit{Cai et al.} \cite{cai2018exploring} trained their models on validation set in addition to training set which increases the effective training set size by 5\%.\par
The focus of this work is not to achieve a state-of-the-art prediction accuracy on a dataset through elaborate selection of network architecture, added robustness through augmentation and etc. The focus of this work is on how we can improve the prediction accuracy given the network architecture and dataset because as years go by better and better CNN architectures are discovered for visual object recognition that can be used for speaker recognition as well. For example, so called Squeeze-and-Excitation Networks \cite{hu2017squeeze} reduced the prediction error of visual object recognition by about 25\%  compared to a ResNet with same architecture without noticeable increase in parameter count. In result, elaborately designed models are likely to be outperformed by more flexible models as larger and larger datasets become available. Furthermore, to show the effectiveness of methods evaluated in this work, we compare our verification accuracies with that of \textit{VoxCeleb2} \cite{Chung18a}, obtained by a much deeper 50 layer ResNet trained on a dataset 5 times larger and much more diverse than \textit{VoxCeleb}.\par

\begin{table*}[t!]\label{comparativeResult}
\centering
\caption{Our models in comparison to other fully supervised trained implementations on \textit{VoxCeleb}. N/R: Not reported}
\resizebox{\linewidth}{!}
{\input{comparativeResults.tex}}
\end{table*}

\section{Conclusions}
\label{sec:conclusions}
In this paper, we investigated different methods that could potentially improve the prediction accuracy of a text-independent speaker recognition system. Results of experiments in this work clearly show that:
\begin{itemize}
    \item Augmenting the training data by repetition and random time-reversion can increase the effective size of training set and thus generalization power of trained network and if this augmentation is applied in testing stage, an improvement in prediction accuracy would be seen.
    \item Similar to \textit{Schroff et al.} \cite{schroff2015facenet}, we confirm that there is an optimal number to dimensionality of speaker embeddings that not only decreases the storage requirement for enrolling new identities but also results in improved verification accuracy.
    \item Use of proposed loss function with independent scale and bias for each class, results in embeddings with superior identification accuracy.
    \item Applying dropout to penultimate fully connected layer can improve the verification accuracy.
\end{itemize}
As the final word, we encourage those interested in implementing speaker recognition systems to apply recommended methods to improve the resulting system's prediction accuracy.
\bibliographystyle{IEEEbib}
\bibliography{main}
\end{document}

%% file: results.tex
  \centering
 \begin{tabular}{cccccccccc}
  \toprule
 & &\multicolumn{4}{c}{without Dropout} &\multicolumn{4}{c}{with Dropout}  \\ \cmidrule(l){2-2}\cmidrule(l){3-6} \cmidrule(l){7-10} 
Loss Function &\#dim  & Top-1 (\%)   & Top-5 (\%)   &  EER (\%) & $C^{min}_{det}$   & Top-1 (\%)   & Top-5 (\%)   &  EER (\%) & $C^{min}_{det}$ \\
  \midrule
Softmax&\begin{tabular}{@{}c@{}} 512 \\256\\128\\64\end{tabular}&
\begin{tabular}{@{}c@{}}89.7 \\89.9\\89.9\\88.6\end{tabular}&
\begin{tabular}{@{}c@{}}96.7 \\ 96.8\\ 96.7 \\96.6\end{tabular}&
\begin{tabular}{@{}c@{}}6.98 \\ 6.78\\ 6.50 \\ 6.27\end{tabular}&
\begin{tabular}{@{}c@{}}0.572 \\ 0.578 \\ 0.556 \\ 0.537\end{tabular}&
\begin{tabular}{@{}c@{}}90.0 \\ 89.5\\ 89.7 \\88.5\end{tabular}&
\begin{tabular}{@{}c@{}}96.6 \\ 96.4\\ 96.4 \\96.0\end{tabular}&
\begin{tabular}{@{}c@{}}6.88 \\ 6.98\\ 6.73 \\ 6.31\end{tabular}&
\begin{tabular}{@{}c@{}}0.540 \\ 0.534 \\ 0.526 \\ 0.527\end{tabular}\\
\midrule
A-Softmax&\begin{tabular}{@{}c@{}} 128\\64\end{tabular}&
\begin{tabular}{@{}c@{}} 56.1 \\ 63.0\end{tabular}&
\begin{tabular}{@{}c@{}} 69.8 \\ 80.9\end{tabular}&
\begin{tabular}{@{}c@{}} 5.63 \\ 4.76\end{tabular}&
\begin{tabular}{@{}c@{}} 0.515 \\ 0.492\end{tabular}&
\begin{tabular}{@{}c@{}} 63.1 \\ 66.7\end{tabular}&
\begin{tabular}{@{}c@{}} 79.0 \\ 83.8\end{tabular}&
\begin{tabular}{@{}c@{}} 4.40 \\ 4.29\end{tabular}&
\begin{tabular}{@{}c@{}} 0.451 \\ 0.442 \end{tabular}\\
\midrule
AM-Softmax&\begin{tabular}{@{}c@{}} 512 \\256\\128\\64\end{tabular} &
\begin{tabular}{@{}c@{}}92.9 \\ 91.4 \\ 92.4 \\ 90.5\end{tabular}&
\begin{tabular}{@{}c@{}}97.6 \\ 96.9 \\ 97.8 \\ 96.3\end{tabular}&
\begin{tabular}{@{}c@{}}5.52 \\ 5.61 \\ 5.46 \\ 5.50\end{tabular}&  
\begin{tabular}{@{}c@{}}0.481 \\ 0.454 \\ 0.476 \\ 0.497 \end{tabular}&
\begin{tabular}{@{}c@{}}93.3 \\ 91.4 \\ 92.8 \\ 90.9\end{tabular}&
\begin{tabular}{@{}c@{}}97.8 \\97.4 \\ 97.5 \\ 97.1\end{tabular}&
\begin{tabular}{@{}c@{}}4.54 \\ 4.52 \\ \bf{4.30} \\ 4.78\end{tabular}&
\begin{tabular}{@{}c@{}}0.432 \\ 0.423 \\ \bf{0.413} \\ 0.417 \end{tabular}\\
\midrule
Logistic Margin&\begin{tabular}{@{}c@{}} 512 \\256\\128\\64\end{tabular}&
\begin{tabular}{@{}c@{}}\bf{94.8} \\ 92.3 \\ 94.6 \\ 92.3\end{tabular}&
\begin{tabular}{@{}c@{}}\bf{98.5} \\ 97.8 \\ 98.1 \\ 97.8\end{tabular}&
\begin{tabular}{@{}c@{}}5.28 \\ 5.44 \\ 4.69 \\ 5.21\end{tabular}&
\begin{tabular}{@{}c@{}}0.469 \\0.485 \\ 0.453 \\ 0.490\end{tabular}&
\begin{tabular}{@{}c@{}}94.3 \\ 91.8 \\ 93.8 \\ 91.8\end{tabular}&
\begin{tabular}{@{}c@{}}98.5 \\ 97.8 \\ 98.3 \\ 97.6\end{tabular}&
\begin{tabular}{@{}c@{}}4.98 \\ 4.96 \\ 4.42 \\ 4.82\end{tabular}&
\begin{tabular}{@{}c@{}}0.453 \\0.472 \\ 0.443\\ 0.483\end{tabular}\\
    \hline
  \end{tabular}

%% file: comparativeResults.tex
  \centering

  \label{tab:comparativeResult}
  \begin{tabular}{cccccccccc}
    \toprule
    Implementation&Dataset(s)&Architecture&Pooling&\#dim  & Top-1 (\%)   & Top-5 (\%)   &  EER (\%) & $C^{min}_{det}$ \\
    \midrule
    VoxCeleb \cite{DBLP:journals/corr/NagraniCZ17}&VoxCeleb1& PLDA+SVM &Variable Length& 200 & 60.8 & 75.6 & 8.8 & 0.73 \\
    VoxCeleb \cite{DBLP:journals/corr/NagraniCZ17}&VoxCeleb1& VGG-M &Variable Length& 1024 & 80.5 & 92.1 & 10.2 & 0.75 \\
    VoxCeleb \cite{DBLP:journals/corr/NagraniCZ17}&VoxCeleb1&VGG-M&Variable Length&256 & N/R & N/R & 7.8 & 0.71 \\
    CNN-TAP \cite{cai2018exploring}& VoxCeleb1 & Thin ResNet-34 & Multi-Crop&128 & 88.5 & 94.9 & N/R & N/R \\
    CNN-TAP \cite{cai2018exploring}& VoxCeleb1 & Thin ResNet-34 & Multi-Crop &128 & N/R  & N/R & 5.27 & 0.439\\
    i-vector \cite{okabe2018attentive}& VoxCeleb1+PRISM & PLDA & Multi-Crop & 400 & N/R & N/R & 5.39 & 0.464\\
    TDNN \cite{okabe2018attentive}& VoxCeleb1+PRISM & TDNN & Multi-Crop & 128 & N/R & N/R & 4.70 & 0.479\\
    LM (ours)  &VoxCeleb1&ResNet-20 & Multi-Crop & 128 & \textbf{94.6} & \textbf{98.1} & \textbf{4.69} & \textbf{0.453} \\
    AMS + dropout (ours)&VoxCeleb1& ResNet-20 & Multi-Crop & 128 &\textbf{ 92.8} & \textbf{97.5 } & \textbf{4.30} & \textbf{0.413} \\
    \midrule
    CNN-LDE \cite{cai2018exploring}& VoxCeleb1 &Thin ResNet-34 & LDE & 128 & 89.9 & 95.7 & N/R & N/R \\
    LDE-ASoftmax \cite{cai2018exploring}& VoxCeleb1 &Thin ResNet-34 & LDE & 128 & N/R & N/R & 4.41 & 0.456 \\
    TDNN \cite{okabe2018attentive}& VoxCeleb1+PRISM & TDNN & Attentive Stat.& 128 & N/R & N/R& 3.85 & 0.406\\
    \midrule
    VoxCeleb2 \cite{Chung18a}&VoxCeleb2& ResNet-50 & Variable Length & 512 & N/R & N/R & 4.19 & 0.449 \\
    VoxCeleb2 \cite{Chung18a}&VoxCeleb2& ResNet-50 & Multi-Crop& 512 & N/R & N/R & 4.43 & 0.454 \\
    VoxCeleb2 \cite{Chung18a}&VoxCeleb2& ResNet-50 & Average Dist. & 512 & N/R & N/R & 3.95 & 0.429 \\
    \bottomrule
  \end{tabular}

%% file: main.bbl
\begin{thebibliography}{10}

\bibitem{DBLP:journals/corr/NagraniCZ17}
Arsha Nagrani, Joon~Son Chung, and Andrew Zisserman,
\newblock ``Voxceleb: a large-scale speaker identification dataset,''
\newblock {\em CoRR}, vol. abs/1706.08612, 2017.

\bibitem{Chung18a}
J.~S. Chung, A.~Nagrani, and A.~Zisserman,
\newblock ``Voxceleb2: Deep speaker recognition,''
\newblock in {\em INTERSPEECH}, 2018.

\bibitem{DBLP:journals/corr/HeZRS15}
Kaiming He, Xiangyu Zhang, Shaoqing Ren, and Jian Sun,
\newblock ``Deep residual learning for image recognition,''
\newblock {\em CoRR}, vol. abs/1512.03385, 2015.

\bibitem{KINNUNEN201012}
Tomi Kinnunen and Haizhou Li,
\newblock ``An overview of text-independent speaker recognition: From features
  to supervectors,''
\newblock {\em Speech Communication}, vol. 52, no. 1, pp. 12 -- 40, 2010.

\bibitem{li2017deep}
Chao Li, Xiaokong Ma, Bing Jiang, Xiangang Li, Xuewei Zhang, Xiao Liu, Ying
  Cao, Ajay Kannan, and Zhenyao Zhu,
\newblock ``Deep speaker: an end-to-end neural speaker embedding system,''
\newblock {\em arXiv preprint arXiv:1705.02304}, 2017.

\bibitem{cai2018exploring}
Weicheng Cai, Jinkun Chen, and Ming Li,
\newblock ``Exploring the encoding layer and loss function in end-to-end
  speaker and language recognition system,''
\newblock {\em arXiv preprint arXiv:1804.05160}, 2018.

\bibitem{liu2017sphereface}
Weiyang Liu, Yandong Wen, Zhiding Yu, Ming Li, Bhiksha Raj, and Le~Song,
\newblock ``Sphereface: Deep hypersphere embedding for face recognition,''
\newblock in {\em The IEEE Conference on Computer Vision and Pattern
  Recognition (CVPR)}, 2017, vol.~1.

\bibitem{srivastava2014dropout}
Nitish Srivastava, Geoffrey Hinton, Alex Krizhevsky, Ilya Sutskever, and Ruslan
  Salakhutdinov,
\newblock ``Dropout: A simple way to prevent neural networks from
  overfitting,''
\newblock {\em The Journal of Machine Learning Research}, vol. 15, no. 1, pp.
  1929--1958, 2014.

\bibitem{okabe2018attentive}
Koji Okabe, Takafumi Koshinaka, and Koichi Shinoda,
\newblock ``Attentive statistics pooling for deep speaker embedding,''
\newblock {\em arXiv preprint arXiv:1803.10963}, 2018.

\bibitem{wang2018additive}
Feng Wang, Jian Cheng, Weiyang Liu, and Haijun Liu,
\newblock ``Additive margin softmax for face verification,''
\newblock {\em IEEE Signal Processing Letters}, vol. 25, no. 7, pp. 926--930,
  2018.

\bibitem{schroff2015facenet}
Florian Schroff, Dmitry Kalenichenko, and James Philbin,
\newblock ``Facenet: A unified embedding for face recognition and clustering,''
\newblock in {\em Proceedings of the IEEE conference on computer vision and
  pattern recognition}, 2015, pp. 815--823.

\bibitem{glorot2010understanding}
Xavier Glorot and Yoshua Bengio,
\newblock ``Understanding the difficulty of training deep feedforward neural
  networks,''
\newblock in {\em Proceedings of the thirteenth international conference on
  artificial intelligence and statistics}, 2010, pp. 249--256.

\bibitem{DBLP:journals/corr/LiMJLZLCKZ17}
Chao Li, Xiaokong Ma, Bing Jiang, Xiangang Li, Xuewei Zhang, Xiao Liu, Ying
  Cao, Ajay Kannan, and Zhenyao Zhu,
\newblock ``Deep speaker: an end-to-end neural speaker embedding system,''
\newblock {\em CoRR}, vol. abs/1705.02304, 2017.

\bibitem{ferrer2011promoting}
Luciana Ferrer, Harry Bratt, Lukas Burget, Honza Cernocky, Ondrej Glembek,
  Martin Graciarena, Aaron Lawson, Yun Lei, Pavel Matejka, Olda Plchot, et~al.,
\newblock ``Promoting robustness for speaker modeling in the community: the
  prism evaluation set,''
\newblock Citeseer.

\bibitem{hu2017squeeze}
Jie Hu, Li~Shen, and Gang Sun,
\newblock ``Squeeze-and-excitation networks,''
\newblock .

\end{thebibliography}
